\documentclass[review]{elsarticle}

\usepackage{amssymb}  
\usepackage{amsthm}
\usepackage{amsmath}
\usepackage{bbm}
\usepackage{multirow}
\usepackage{algorithm} 
\usepackage{algpseudocode}

\usepackage{hyperref}
\journal{Journal of \LaTeX\ Templates}

\biboptions{authoryear}

\begin{document}

\begin{frontmatter}

\title{A differential neural network learns stochastic differential equations and the Black-Scholes equation for pricing multi-asset options}

\author{Sang-Mun Chi}
\address{Department of Computer Science, Kyungsung University, Busan 48434, South Korea}

\begin{abstract}
Neural networks with sufficiently smooth activation functions can approximate values and derivatives of any smooth function, and they are differentiable themselves.
We improve the approximation capability of neural networks by utilizing the differentiability of neural networks; 
the gradient and Hessian of neural networks are used to train the neural networks to satisfy the differential equations of the problems of interest.
Several activation functions are also compared in term of effective differentiation of neural networks.
We apply the differential neural networks to the pricing of financial options,
where stochastic differential equations and the Black-Scholes partial differential equation 
represent the relation of price of option and underlying assets,
and the first and second derivatives, Greeks, of option play important roles in financial engineering.
The proposed neural network learns --
(a) the sample paths of option prices generated by stochastic differential equations 
and (b) the Black-Scholes equation at each time and asset price.
Option pricing experiments were performed on multi-asset options such as exchange and basket options.
Experimental results show that the proposed method gives accurate option values and Greeks;
sufficiently smooth activation functions and the constraint of Black-Scholes equation 
contribute significantly for accurate option pricing.
\end{abstract}


\end{frontmatter}

%\linenumbers

\section{Introduction}

% differential neural networks
Neural networks (NNs) have found successful applications such as
image understanding, handwriting recognition, speech recognition, and drug discovery
\cite{hinton, ng, jugen, deng, Ramsundar}.
This success is due to the universal approximation capability of NNs,
approximating the function accurately without assuming any mathematical relationship between the input and the output of a function.
Whereas no mathematical relationships are found for the foregoing applications,
strong governing differential equations are presented in many fields such as finance, fluid dynamics, quantum mechanics, and diffusion phenomena.
For these problems, the approximation capabilities of NNs can be improved
when the differential relationships are included in the modeling.

Neural networks with sufficiently smooth activation functions can approximate a function and its derivatives \cite{Cybenko, HORNIK1990, GALLANT1992129};
they are also differentiable because the composition of differentiable functions is differentiable.
Thus, we can obtain the gradient and Hessian of differentiable NNs, and use these derivatives to make the NNs themselves satisfy the differential equations of the problems of interest.
To obtain accurate derivatives of NNs, 
we calculate the gradient and Hessian of NNs using automatic differentiation \cite{Nocedal, tf, raissi2018}, 
which is an exact method based on the chain rule for differentiating compositions of functions.
% activation function, Greeks
Since the differentiation of NNs plays a key role in the present work, 
activation functions should be sufficiently smooth as well as effective for our applications.
Although any smooth activation function makes NNs differentiable, 
we need the activation functions which show good practical performance; we compare several activation functions in terms of estimation accuracy for the value, gradient, and Hessian.

% benchmark on option pricing
We need suitable benchmark problems to examine the effectiveness of our differential NN.
We consider financial option pricing because the option pricing problem has a widely known mathematical relationship, the Black-Scholes equation \cite{Black1973}, and the derivatives of option price, Greeks, are extensively studied since they have many practical applications such as hedging or speculating the future asset price \cite{Hull, Shreve2004, glasserman}.
Hence, the performance of the proposed differential NNs can be easily tested
using the experiments on option pricing. 

% comparing FDM, MC
Most multidimensional options have no general closed-form solution for pricing
and they are priced by numerical approximation techniques.
Various numerical methods have therefore been developed to solve this problem such as  finite differences methods (FDMs), Fourier methods, and Monte Carlo (MC) simulations \cite{Hull, Shreve2004,  Carr1999, glasserman}. 
FDMs approximate the discrete version of Black-Scholes partial differential equation (PDE);
MC methods use the generated sequence of stock prices governed by the stochastic differential equation (SDE).
The present work combines the information of the PDE and SDE,
and trains NNs using --
(a) the option price generated by the SDE and (b) the constraint of the Black-Scholes equation.
As a result, the proposed method becomes mesh-free method like MC simulations, and does not suffer from the curse of dimensionality associated with high-dimensional FDMs.
Furthermore, Greeks are easily computed by differentiating the neural network
because the proposed method represents the solution of Black-Scholes equation in the form of
a neural network.

% NN option pricing% comparing NN options, SD+BS learning,
NNs have also been applied for option pricing problems
\cite{Gencay2001, Kohler2010, Sirignano2018,Wei2017, Becker2019}.
Our approach relates to methods of differentiating NNs used in \cite{Sirignano2018, raissi2018}.
We use automatic differentiation method to compute the gradient and Hessian of NNs, 
while \cite{Sirignano2018} use a MC method to approximate the Hessian;
automatic differentiation is an exact operation and not a kind of approximation methods used in MC methods and FDMs \cite{Nocedal, tf}.
The present work generates training data using the SDE which gives more realistic evolution of asset prices 
while  \cite{Sirignano2018}  use random samples from the region where the function is defined.
\cite{raissi2018} uses automatic differentiation method to compute the gradient of NNs
and SDE to generate training data;
but, the loss function for training NN does not contain PDE and the Hessian of NNs are not used.
In addition, we adopt sufficiently smooth activation function softplus \cite{softplus} for the efficient differentiation of NNs based on the performance comparison experiments, 
while \cite{Sirignano2018} use tanh and \cite{raissi2018} uses sine as an activation function.

The remainder of the paper is organized as follows.
We discuss the related works on multidimensional option pricing and numerical approximation methods in Section 2.
The proposed method is developed for learning stochastic differential equations and the Black-Scholes equation for option pricing in Section 3.
The validity of the proposed method is demonstrated in Section 4, and conclusions are drawn in Section 5.

\section{Price and Greeks of multidimensional options}

A financial option is a contract between the seller and the buyer (holder of the option).
A European call option gives the holder the right but no obligation 
to buy the risky assets at expiration time $T$ for a price that is agreed on now, the strike price $K$.
The option itself has a price because it give the holder a right.
Let the stock price $S_i(t)$ be a geometric Brownian motion, 
\begin{equation}
dS_i(t)= r S_i(t) dt + \sigma_i S_i(t) d\tilde W_i(t), \label{eq:sde}
\end{equation}
where $t$ denotes time, $r$ risk-free interest rate,
$\sigma_i$ the volatility of the asset price, and $\tilde W_i$ a Brownian motion under a risk-neutral probability measure with 
covariance $Cov(\tilde W_i (t), \tilde W_j (t)) = \rho_{ij} t$.
The correlated Brownian motion $\tilde W_i (t)$ can be expressed as $\sqrt{t} L_i Z$ 
where $Z$ is a standard $n$-dimensional normally distributed vector
and $L_i$ is the $i$th row of $L$ ;
$L$ represents any matrix such that $\Sigma = LL^\prime$ where the $(i,j)$ component of $\Sigma$ is $\rho_{ij}$.
We choose $L$ to be the Cholesky factor of  $\Sigma$.
Then the stock prices between time $t_1$ and $t_2$ are related by
\begin{equation}
S_i(t_2)= S_i(t_1)  \exp ((r-\frac{1}{2} \sigma_i^2)(t_2 - t_1)+ \sigma_i \sqrt{t_2 - t_1} L_i Z)\label{eq:stock}
\end{equation}
\cite{glasserman, Shreve2004, Hull}.

\subsection{Exchange option}\label{sec:exchange}
An exchange option gives the holder the right to exchange one asset for another in given time
in the future and it is commonly seen in the energy market.
The payoff of exchange option is
\begin{equation}
	\max(S_1(T) - S_2(T), 0).\label{eq:exop}
\end{equation}
\cite{Margrabe1978} developed an analytical solution formula
for the price of this option.
The price of exchange option at time $t$ is given by
\begin{equation}
u(t,S_1, S_2) = S_1 N (d_1) - S_2 N(d_2), \label{exCall}
\end{equation}
where
$ \sigma = \sqrt{ \sigma_1^2 + \sigma_2^2 - 2 \rho_{12} \sigma_1 \sigma_2},
\quad d_1= \frac{1}{\sigma \sqrt{T-t}} \left[\ln{\left(\frac{S_1}{S_2}\right)} + \frac{\sigma^2}{2} (T-t)  \right],
\quad d_2= d_1 - \sigma \sqrt{T-t}$,
and $N(x)=\frac{1}{\sqrt{2\pi}} \int_{-\infty}^{x} \mathrm e^{-\frac{1}{2}z^2} dz. $

Greeks are the sensitivities of the option price to the movement of various parameters.
They are used for risk management;
the risk in a short position in an option is offset by holding delta units of each underlying asset, where the delta is the partial derivative of the option price with respect to the current price of
that underlying asset. Sensitivities with respect to other parameters are also
widely used to measure and manage risk.
The Greeks of exchange option is obtained by differentiating Eq.\eqref{exCall}.
Let $\phi (x) = \frac{ 1}{\sqrt{2\pi} } \exp(-\frac{ x^2}{2})$,
differentiating Eq.\eqref{exCall} by $S_1$ and using the property $\phi (d_1) / \phi (d_2) = S_2 / S_1$
gives the delta
\begin{equation}
\Delta = \frac{\partial u}{\partial S_1} = N(d_1).\label{exdelta}
\end{equation} 
In some other hedging strategy, we need to hedge away the risk due to the
changes of underlying asset's delta; the gamma is defined by
\begin{equation}
\Gamma = \frac{\partial \Delta}{\partial S_1} = \frac{\partial^2 u}{\partial S_1^2}
%= \frac{\partial N(d_1)}{\partial S_1}
%= \frac{\partial d_1}{\partial S_1} N' (d_1)
= \frac{ \phi (d_1)}{S_1 \sigma \sqrt{T-t} }.\label{exgamma}
\end{equation}
The time decay of the value for an option is called theta, given by
\begin{equation}
\Theta = \frac{\partial u}{\partial t}
=-\frac{\sigma}{4\sqrt{T-t} }\left[S_1 \phi (d_1) + S_2 \phi (d_2)\right].\label{extheta}
\end{equation}

\subsection{Basket option}
A European basket call option gives the holder to buy a group of underlying assets
at the same time.
The price of the option at maturity are given by
\begin{equation}
C(T,S_1(T), S_2(T), \ldots, S_n(T)) = \max( \sum_{i=1}^{n} w_i S_i(T) - K,0)\label{eq:BasCall}
\end{equation}
where $w_i$ denotes the quantity of $i$th asset in basket option contracts.
Since this option has no exact formula for price and Greeks,
these values are approximated by numerical methods such as
finite difference methods (FDMs) and Monte Carlo (MC) simulations \cite{Hull, Shreve2004, glasserman}.
While FDMs are accurate in low dimensional problems, they become infeasible in higher dimensions due to increased number of grid points and numerical instability.
MC methods are versatile for handling general type of options, multi-asset, and path dependent problems.
But, the MC method converges slow at rate $O(1/\sqrt{M})$, where $M$ denotes the number of random samples.

% FDM
FDMs approximate the price of option, $u(t,S_1,S_2, \ldots, S_n)$, on the basis of the Black-Scholes equation \cite{Hull, Shreve2004},
\begin{equation}
	\frac{\partial u}{ \partial t } + \frac{1}{2} \sum_{i=1}^{n} \sum_{j=1}^{n} 
\sigma_i \sigma_j \rho_{i,j}  S_i S_j \frac{\partial^{2} u}{\partial S_i \partial S_j}
	+ r  \sum_{i=1}^{n} S_i \frac{\partial u}{\partial S_i} =	r u, \label{eq:mBS}
\end{equation}
where $r$ denotes risk-free interest rate,
and $(\sigma_i \sigma_j \rho_{i,j})_{1 \le i,j \le n}$ denotes the covariance matrix of the stock prices.
FDMs discretize the stock price and time dimensions 
and use the option values at the time of expiration.
We use central difference equation to obtain the first and second derivatives with respect to stock price, and forward difference equation to obtain time derivatives.
The Greeks at time zero are obtained using the solved $u(\cdot)$.
The delta is given by
\begin{equation}
\Delta_i = \frac{u(0,S_1, \ldots, S_i + \delta_s, \ldots, S_n)-u(0,S_1, \ldots, S_i - \delta_s, \ldots, S_n)}{2\delta_s},
\end{equation} 
where $\delta_s$ denotes a stock price discretization unit.
The gamma is given by
\begin{equation}
\Gamma_i = \frac{u(0,\ldots, S_i + \delta_s, \ldots)-2u(0,\ldots, S_i , \ldots)+u(0,\ldots, S_i -\delta_s, \ldots)}{\delta_s^2}
\end{equation}
The theta is given by
\begin{equation}
\Theta =  \frac{u(\delta_t,S_1, \ldots, S_n)-u(0,S_1, \ldots, S_n)}{\delta_t},
\end{equation} 
where $\delta_t$ denotes a time discretization unit.

\begin{table}[t]
\centering
\begin{tabular}{|c|c|}
\hline
 Greeks & Expectation of the following formulas\\ \hline
 \multicolumn{2}{c}{Exchange option} \\ \hline
$\Delta$ & $e^{-rT} \frac{S_1(T)}{S_1(0)}  \mathbbm{1} \{S_1(T) > S_2(T) \} $  \\
\hline
$\Theta$ & $e^{-rT} [ S_1(T) ( \frac{\sigma_1^2}{2} - \frac{\sigma_1 L_1 Z}{2 \sqrt{T} } ) 
- S_2(T) ( \frac{\sigma_2^2}{2} - \frac{\sigma_2 L_2 Z}{2 \sqrt{T} } ) ] \mathbbm{1} \{S_1(T) > S_2(T) \}$  \\
\hline
$\Gamma$ & $e^{-rT}  \frac{ ( Z^{\prime} L^{-1} A^{-1} )_1 }{\sqrt{T} }  \frac{ S_2 (T) }{ S_1^2(0)}
 \mathbbm{1} \{(S_1(T) > S_2(T) \} $  \\ \hline
 
 \multicolumn{2}{c}{Basket option} \\ \hline
$\Delta_i$ & $e^{-rT} \frac{S_i(T)}{S_i(0)}  w_i \mathbbm{1} \{ \sum_{i=1}^{n} w_i S_i(T) > K \}$  \\ \hline
$\Theta$ & $ e^{-rT} [  -rK + \sum_{i=1}^{n} w_i S_i(T) ( \frac{\sigma_i^2}{2} - \frac{\sigma_i L_i Z}{2 \sqrt{T} } ) ]
 \mathbbm{1} \{ \sum_{i=1}^{n} w_i S_i(T) > K \} $  \\ \hline
$\Gamma_i$ & $e^{-rT}  \frac{ ( Z^{\prime} L^{-1} A^{-1} )_i }{\sqrt{T} S_i^2(0) }   
\{ w_i S_i(T) - \sum_{i=1}^{n} w_i S_i(T) + K \}$ \\
&  $ \times \mathbbm{1} \{ \sum_{i=1}^{n} w_i S_i(T) > K \} $  \\ \hline
\end{tabular}
\caption{Greeks estimation by MC simulation.}
\label{tab:Greeks}
\end{table}

% MC
MC methods generate multiple simulated paths of stock price by Eq.~\eqref{eq:stock}.
The price of options is obtained by averaging the payoff as follows.
\begin{equation}
V(t) =  E_Q [ e^{-r(T-t)}V(T) | \mathcal F(t)], \label{eq:mcPrice}
\end{equation} 
where $V(T)$ is the payoff function,
$\mathcal F(t)$ is a filteration for the Brownian motion $\tilde W_i (t)$ 
in Eq.\eqref {eq:sde} and $Q$ is the risk-neutral measure \cite{Hull, Shreve2004, glasserman}.
The Greeks estimation is a practical challenge in MC methods.
In the present work, the pathwise derivative estimate is used for calculating the delta and theta, 
and combination of likelihood ratio method and pathwise derivative estimate is used for calculating gamma.
The Greeks at time zero are given by averaging the formulas in Table ~\ref{tab:Greeks}.
Derivations of these formulas are described in Appendix.

\section{Learning stochastic differential equations and the Black-Scholes equation}

\subsection{Learning stochastic differential equations}
We approximate the price of option $u(t,S_1,S_2, \ldots, S_n)$ in Eq.\eqref{eq:mBS}
using a neural network.
Expiry time $T$ is divided into $N(=200)$ equal intervals $t_0=0 < t_1 < t_2 < \cdots < t_N=T$.
Multiple stock price $S(t_k) =(S_1 (t_k), S_2 (t_k), \ldots, S_n (t_k)) $ is generated by Eq.~\eqref{eq:stock}
\begin{equation}
S_i(t_k)= S_i(t_{k-1})  \exp ((r-\frac{1}{2} \sigma_i^2)(t_k - t_{k-1}) + \sigma_i d \tilde W_i(t_k))\label{eq:istock}
\end{equation}
with an initial stock price $S(t_0)=S(0)$ and $d \tilde W_i(t_k) = \sqrt{t_k - t_{k-1}} L_i Z\label{eq:dw}$.
The price of option $u(t_k,S(t_k))$ is approximated by the neural network $\mathcal N(t, S)$
\begin{align}
& \mathcal N^1 (t_k,S(t_k)) = f [ w^1 (t_k, S(t_k))  + b^1 ], \nonumber \\
& \mathcal N^2 (t_k,S(t_k)) = f [ w^2 \mathcal N^1 (t_k,S(t_k))  + b^2 ], \nonumber \\
& \cdots \label{eq:C}, \\
& \mathcal N(t_k,S(t_k)) = \mathcal N^l (t_k,S(t_k)) =  w^l \mathcal N^{l-1} (t_k,S(t_k)) + b^l , \nonumber
\end{align}
where $f$ is an activation function, $w^l$ and $b^l$ parameters of $l$th layer of the neural network. In the present work, the number of layers $l$ is five,  the number nodes 35 for layer $l=1,2,3,4$, and one in the last layer $l=5$.

To train the neural network $\mathcal N(t, S)$, we need to know the target $u(t, S)$.
Although the $u(\cdot)$ is of course unknown at $t<T$, 
this function satisfies the following SDE when the stock price follows Eq.\eqref{eq:sde}  \cite{Hull, Shreve2004, glasserman}.
\begin{equation}
du = ( u_t + \sum_{i=1}^{n} r S_i  u_{S_i}  +  \frac{1}{2} S^\prime H S  )dt + 
\sum_{i=1}^{n} \sigma_i S_i  u_{S_i} d\tilde W_i, \label{eq:uSDE}
\end{equation}
where $u_t = \frac{\partial u}{\partial t}, S^\prime = (S_1,S_2, \ldots, S_n), H_{ij} = \rho_{ij}\sigma_i \sigma_j \frac{\partial^2 u}{\partial S_i \partial S_j}$
and $u_{S_i}= \frac{\partial u}{\partial S_i}$.
Based on this SDE and the derivatives of the neural network $\mathcal N(t, S)$,
we generate the following $\tilde u(t_k, S(t_k))$ that approximates $u(t_k, S(t_k))$.
 
\begin{align}
\tilde u(t_k, S(t_k)) &= \mathcal N(t_{k-1}, S(t_{k-1}))  + \{ \mathcal N_t(t_{k-1}, S(t_{k-1})) + \nonumber \\ 
& \sum_{i=1}^{n} r S_i(t_{k-1}) \mathcal N_{S_i}(t_{k-1}, S_i(t_{k-1}))  +  \frac{1}{2} S(t_{k-1})^\prime H_{k-1} S(t_{k-1})  \} dt_k \nonumber \\
& + \sum_{i=1}^{n} \sigma_i S_i(t_{k-1}) \mathcal N_{S_i}(t_{k-1}, S_i(t_{k-1})) d\tilde W_i (t_k),\label{eq:Ctilde}
\end{align}
where  $\mathcal N_t$ and $\mathcal N_{S_i}$ respectively denote partial derivatives of the neural network $\mathcal N(t, S)$ with respect to $t$ and $S_i$,
$(H_{k-1})_{ij} = \rho_{ij}\sigma_i \sigma_j \frac{\partial^2 \mathcal N(t_{k-1}, S(t_{k-1}) )}{\partial S_i \partial S_j}$, $dt_k = t_k - t_{k-1}$ and 
$d \tilde W_i(t_k) = \sqrt{t_k - t_{k-1}} L_i Z$.
This $\tilde u(t_k, S(t_k))$ is used to train $\mathcal N(t_k, S(t_k))$ 
with the following loss function which is minimized during training the neural network.
\begin{equation}
L_{SDE} = \sum_{k=1}^N [ \tilde u(t_k, S(t_k)) - \mathcal N(t_k, S(t_k))  ]^2.\label{eq:lossC}
\end{equation}

\subsection{Learning the Black-Scholes equation}\label{sec:Loss}
We enforce the neural network $\mathcal N(t, S)$ to
satisfy the Black-Scholes equation in Eq.~\eqref{eq:mBS}
by minimizing the following representation of Black-Scholes equation in the form of a neural network
\begin{align}
L_{BS} = & \sum_{k=1}^m [ \mathcal N_t(t_k, S(t_k))  + \sum_{i=1}^{n} r S_i(t_k) \mathcal N_{S_i}(t_k, S_i(t_k)) \nonumber  \\
& +  \frac{1}{2} S(t_k)^\prime H_k S(t_k)  - r \mathcal N(t_k, S(t_k))  ]^2.\label{eq:lossD}
\end{align}

We need to differentiate the neural network with respect to variables $t_k$ and $S(t_k)$
for the calculation of Eq.~\eqref{eq:Ctilde} and Eq.~\eqref{eq:lossD}.
This differentiation is performed by using automatic differentiation. 
Automatic differentiation is an exact differentiation method, not an approximation method,
so that it gives accurate differential results.
It apply the chain rule for differentiating compositions of a set of elementary functions for which derivatives
are known exactly \cite{Nocedal, tf, raissi2018}.
Since software libraries like Tensorflow \cite{tf} already provide operations for automatic differentiation,
we used the tensorflow.GradientTape operation in TensorFlow to calculate the gradient of neural network; tensorflow.GradientTape operation is performed to the gradient of NN for the calculation of the Hessian of NN.

At expiration time $T$, true $u(T,S_1,S_2, \ldots, S_n)$ is given by the payoff function.
To match the option value at maturity time $T$, the third loss function is
\begin{equation}
L_T =  [ \mathcal N(T, S(T)) - h(T)  ]^2 * w_T,\label{eq:lossT}
\end{equation}
where $h(T)$ denotes the payoff function such as 
Eq.\eqref{eq:exop} and Eq.\eqref{eq:BasCall},
and $w_T$ denotes the weight for accurate approximation at $T$, $w_T = N / 20 (=10.)$ as a default weight in this work.
Finally, we use the following loss function that penalizes 
the neural network output deviations from the stochastic differential equation,
the neural network differential structure deviations from Black-Scholes equation,
and the neural network output deviations from the payoff function at expiration time $T$,
\begin{equation}
Loss = L_{SDE} + L_{BS} + L_T\label{eq:loss}.
\end{equation}

\subsection{Training and estimation procedures of the neural network}

We call our neural network SDBS, because it is based on the stochastic differential equation and the Black-Scholes equation. We summarize the training procedure of SDBS in Algorithm 1.
For updating the parameters of SDBS (line 12 in Algorithm 1),
we use a stochastic gradient descent method, adam, that is based on adaptive estimation of first-order and second-order moments \cite{adam}.
It is commonly observed that a monotonically decreasing learning rate results in a better performing model.
The present work uses a PolynomialDecay schedule in Tensorflow and the learning rate is linearly decayed from
$10^{-3}$ to $10^{-7}$. As the final trained model, we choose the model which has the minimum $Loss$ during training iterations.

\begin{algorithm}
	\caption{Training of SDBS.}
	
	\begin{algorithmic}[1]
	 	\State \textbf{Input: }  $S(t_0)$, $T, \sigma_i, r, \rho_{ij}, N$, nEpoch
	 	\State Calculate $L$ representing $\Sigma = LL^\prime$, $t_0 = 0$
		\For {epoch=1, 2, $\cdots$, nEpoch}
		\State Calculate $\mathcal N, \mathcal N_t, \mathcal N_{S_i}, H$ at $(t_0, S(t_0))$ 
		
		\For {$k=1, 2, \cdots, N$}
		\State $t_k = t_{k-1} + T/N$
		\State Calculate $\tilde u(t_k, S(t_k))$ by Eq.~\eqref{eq:Ctilde}
		\State Calculate $S(t_k)$ by Eq.~\eqref{eq:istock} and   
		\State Calculate $\mathcal N, \mathcal N_t, \mathcal N_{S_i}, H$ at $(t_k, S(t_k))$ 
		\EndFor
		
		\State Calculate $Loss$ by Eq.~\eqref{eq:loss}
		\State Update the parameters of neural network by minimizing $Loss$
		 		
		\If {$Loss$ is reduced comparing with that in the previous epoch}
		\State Save parameters of SDBS  
		\EndIf		

		\EndFor	
	\end{algorithmic} 
\end{algorithm} 

The estimation procedure is similar to the training procedure.
There are three differences:
(1) in addition to the input of training procedure, the parameters of trained SDBS is loaded,
(2) the learning rate is fixed by $10^{-7}$ in line 12 of Algorithm 1,
(3) the line 13, 14, and 15 are replace by output Price $= \mathcal N(t,S(t) )$, $\Delta_i = \mathcal N_{S_i}(t,S(t) )$,
$\Gamma_i =  \mathcal N_{S_i S_i}(t,S(t) )$, $\Theta = \mathcal N_t(t,S(t) )$.

\subsection{Activation functions for differentiable SDBS}\label{sec:mathActi}

\begin{table}[t]
\centering
\begin{tabular}{|l|c|c|c|}
\hline
Name & Equation & Range & Order of continuity\\
\hline
sigmoid & $1/(1+e^{-x})$ & (0, 1) & $C^\infty$\\
tanh & $(e^x - e^{-x}) / (e^x+e^{-x})$ & (-1, 1) & $C^\infty$\\
sin & $\sin(\cdot)$ & (-1, 1)  & $C^\infty$\\
relu & 
  $ \left\{
  \begin{array}{rl}
  0 & \text{if } x \le 0,\\
  x & \text{if } x > 0.
  \end{array} \right. $
 & $(0, \infty)$  & $C^0$\\
elu &  
  $ \left\{
  \begin{array}{rl}
  \alpha (e^x - 1) & \text{if } x \le 0,\\
  x & \text{if } x > 0.
  \end{array} \right. $
 & $(-\alpha, \infty)$  
 & $ \left\{
  \begin{array}{rl}
  C^1 & \text{if } \alpha = 1,\\
  C^0 & \text{otherwise} .
  \end{array} \right.$ \\
selu &  
  $ \lambda \left\{
  \begin{array}{rl}
  \alpha (e^x - 1) & \text{if } x \le 0,\\
  x & \text{if } x > 0.
  \end{array} \right.$
 & $(-\lambda \alpha, \infty)$ 
 & $ C^0$ \\
softplus & $\ln (1+e^x)$ & $(0, \infty)$ & $C^\infty$\\
\hline
\end{tabular}
\caption{Classification of activation functions}
\label{tab:activation}
\end{table}

Activation functions determine the output of a node and the characteristics of NNs.
They have various shape, range, order of continuity, and magnitude of gradient. 
Every function in Table ~\ref{tab:activation} has nonlinear shape
because nonlinearity of the activation functions allows NNs to be universal function approximators \cite{Cybenko, HORNIK1990}.
The range of function is finite for sigmoid, tanh, and sin while infinite for relu \cite{relu}, elu \cite{elu}, selu \cite{selu}, softplus \cite{softplus}.
The boundedness and continuity of partial derivatives of the activation functions up to order $m$ are required to approximate the functions with all partial derivatives up to $m$ are continuous and bounded \cite{HORNIK1990, GALLANT1992129}.
Hence, the order of continuity of activation functions critically affects the performance of the proposed method
in which the first and second derivatives of a neural network play a key role.
This consideration suggests that the following sufficiently smooth activation functions are suitable for the present work.
\begin{equation}
C^\infty \ \text{functions} = \{ \text{sigmoid, tanh, sin, softplus} \} \label{eq:smoothAct}.
\end{equation}
The magnitude of gradient of activation function should not vanish for the update of parameters of NNs during training.
When the sigmoid and tanh are either too high or too low, the gradient vanishing is observed.
The following relu-like functions have been widely used in many neural network applications 
because they show high performance and their gradients are non-zero for all positive values.
\begin{equation}
\text{relu-like functions} = \{ \text{relu, elu, selu, softplus} \}, \label{eq:reluAct}
\end{equation}
where $\alpha=1$ is used for elu 
and pre-defined constants $\alpha=1.67326324$ and $\lambda=1.05070098$ in Tensorflow are used for selu.

In section \ref{sec:activation}, we compare the performance of  activation functions in Table ~\ref{tab:activation} 
in terms of estimation accuracy for value, the first and second derivatives of NNs.

\section{Numerical experiments}

In this section, numerical experiments are performed to demonstrate the effectiveness of the proposed method for pricing multidimensional options such as exchange and basket options.

\subsection{Estimation of price and Greeks of exchange options }\label{sec:exNu}

We can compare the performance of numerical methods using an exchange option 
because this option has an exact solution.
\cite{Margrabe1978} provided the formula for the exchange option price in Eq.\eqref{exCall}.
By differentiating the Margrabe formula, we obtained the Greeks in Eq.\eqref{exdelta}-\eqref{extheta}.
In the same notation in section \ref{sec:exchange},
initial stock prices $S_1 (0) = 60$ and $S_2(0) = 60$,
time to maturity $T=1$, 
volatility of first asset $\sigma_1 = 0.4$,
volatility of second asset $\sigma_2 = 0.2$,
interest rate $r=0.1$, 
and correlation coefficient $\rho_{12}=0.4$ are parameters of the first experiment.
We define the error between exact solution and the estimate by relative error 
\begin{equation}
\text{ rError} = \left | \frac{exact - estimate}{exact} \right |.\label{eq:error}
\end{equation}

\begin{table}
\centering
\small
\begin{tabular}{|c|c|c|c|c|c|c|}
\hline
 & Exact & FDM1 & FDM2 & MC1 & MC2 &MC3  \\
\hline
Price & 8.777591 & 8.765359 &  8.776234 & 8.784203 &8.776402 &8.777109  \\
rError & 0 & 1.39e-03  &  1.55e-04  & 7.53e-04 & 1.35e-04  & \textbf{5.49e-05} \\
\hline
$\Delta$ & 0.573140 &0.572740    & 0.573102  & 0.573611 & 0.573094 & 0.57313 \\
rError & 0 & 7.09e-04 &7.80e-05 & 8.10e-04 &9.24e-05 & \textbf{2.97e-05} \\
\hline
$\Gamma$ &0.017726& 0.017728 & 0.017726&0.017738&0.017724 & 0.017725\\
rError & 0 & 1.15e-04 &\textbf{1.07e-05} &6.82e-04&9.47e-05 & 4.06e-05 \\
\hline
$\Theta$ & -4.339281 &-4.344155&-4.339812&-4.343678&-4.338946  & -4.339065\\
rError &0 & 1.12e-03 &1.22e-04 & 1.01e-03 &7.72e-05 & \textbf{4.99e-05} \\
\hline
\end{tabular}
\caption{Numerical estimate of the price and Greeks of the exchange option. The best results are highlighted in bold face.}
\label{tab:exRes}
\end{table}

\begin{table}
\centering
\small
\begin{tabular}{|c|c|c|c|c|c|c|}
\hline
nEpoch & Exact & 25,000 & 50,000 & 100,000 &200,000 & 400,000\\
\hline
Price & 8.777591 &8.777736 & 8.777681 &8.777660 &8.777587 & 8.777516  \\
rError & 0 & 1.65e-05 &  1.02e-05 & 7.85e-06 & \textbf{4.85e-07} & 8.51e-06 \\
\hline
$\Delta$ & 0.573140 &0.573155  &0.573158   &0.573152  & 0.573139 & 0.573152 \\
rError & 0 & 1.40e-05 &1.84e-05 & 9.44e-06 &1.45e-05 & \textbf{9.37e-06} \\ 
\hline
$\Gamma$ &0.017726&0.017692  &0.017653 &0.017699 &0.017708 &0.017701 \\
rError & 0 & 1.90e-03 &4.12e-03&1.52e-03&\textbf{1.01e-03} & 1.38e-03 \\
\hline
$\Theta$ & -4.339281 &-4.323961 &-4.311028 &-4.330462&-4.330742 &-4.331632  \\
rError & 0 & 3.53e-03  &6.51e-03 & 2.03e-03 &1.97e-03 & \textbf{1.76e-03} \\
\hline
\end{tabular}
\caption{SDBS estimate of the price and Greeks of the exchange option using several training epochs. The best results are highlighted in bold face.}
\label{tab:exSDBS}
\end{table}

In Table ~\ref{tab:exRes}, FDMs discretize the domain of stock price $[0, 300]$ with 100 uniform intervals for FDM1 and 300 for FDM2,
and the time domain $[0, T]$ with 5,000 uniform intervals  for FDM1 and 50,000 for FDM2.
MC simulations calculate the price and Greeks of options given by
Eq. \eqref{eq:mcPrice} and Table \ref{tab:Greeks}.
MC1 uses $10^7$ simulated paths, $10^{8}$ for MC2,  and $10^{9}$ for MC3.
As can be seen in Table ~\ref{tab:exRes}, FDMs and MC methods give accurate estimation.
Furthermore, the finer discretization makes FDMs more accurate; MC methods estimate more accurately with more simulated paths.

Five independent SDBS models are trained for several nEpochs using the Algorithm 1.
The batch size 10,000 of simulated stock price $S(t)$ was used for training and estimation procedure.
The price and Greeks are generated 1000 times for each trained NN.
The average of 5,000 (5 models x 1,000 times) values is used as the final estimated values.
Comparing with the results in Table ~\ref{tab:exRes}, the SDBS in Table \ref{tab:exSDBS} gives comparable performance to the classical FDM and MC methods. In particular, the estimation of price for nEpoch 200,000 is significantly accurate.
These experiments show that the differential neural network based SDBS performs accurately 
which estimates Price $= \mathcal N(t_0,S(t_0) )$, $\Delta_i = \mathcal N_{S_i}(t_0,S(t_0) )$,
$\Gamma_i =  \mathcal N_{S_i S_i}(t_0,S(t_0) )$, and $\Theta = \mathcal N_t(t_0,S(t_0) )$.

\begin{table}[t]
\centering
\begin{tabular}{|c|c|c|c|c|c|c|}
\hline
  \multicolumn{2}{|c|}{$S_1(0), S_2(0)$} & 20,60 & 40,60 & 60,60 & 60,40 & 60,20 \\
\hline
\multirow{3}*{ Price} &FDM2&5.64e-02 &5.55e-04 & 1.55e-04 & \textbf{4.08e-07} &\textbf{1.81e-08}     \\
& MC2 & \textbf{3.13e-03} & \textbf{2.36e-05} & 1.35e-04 &9.42e-05 & 6.75e-05   \\
& SDBS & 1.25e-02 &6.22e-05 & \textbf{7.85e-06} &1.31e-06 &  2.90e-07 \\
\hline
\multirow{3}*{$\Delta$ } &FDM2  &4.67e-02 &2.15e-04 & 7.80e-05&1.03e-04 &3.28e-06 \\
& MC2 &\textbf{1.52e-03} & 2.91e-04& 9.24e-05&1.20e-04 &4.21e-05\\
& SDBS &2.22e-03 & \textbf{6.63e-05} & \textbf{9.44e-06} & \textbf{1.78e-05} & \textbf{2.25e-06} \\
\hline
\multirow{3}*{$\Gamma$} & FDM2  &4.06e-02 & \textbf{1.70e-04} & \textbf{1.07e-05} & 2.48e-04 & \textbf{2.07e-04}\\
& MC2  &1.74e-03  & 2.55e-04& 9.47e-05& \textbf{5.14e-05} & 2.06e-02\\
& SDBS &\textbf{1.36e-03} & 2.11e-03 &1.52e-03 & 4.14e-03 &8.42e-02 \\
\hline
\multirow{3}*{$\Theta$} & FDM2 &8.95e-02 & \textbf{6.51e-06} & 1.22e-04& \textbf{2.82e-05} & \textbf{2.07e-04}\\
& MC2 &1.15e-03 &1.55e-04 & \textbf{7.72e-05} & 2.71e-04&2.39e-02\\
& SDBS & \textbf{8.80e-04} & 3.73e-03 &2.03e-03 &6.84e-03 &1.09e-01 \\
\hline
\end{tabular}
\caption{rErrors of numerical methods for the exchange options with several initial stock prices. The best results are highlighted in bold face. }
\label{tab:exSDBSs1s2}
\end{table}

Table \ref{tab:exSDBSs1s2} shows the estimates of price and Greeks of exchange options with  different initial stock price pairs ($S_1 (0)$, $S_2(0)$), where the SDBS uses the same experimental configuration as in Table \ref{tab:exSDBS} with 100,000 nEpoch.
In the estimation of Price and $\Delta$, the SDBS shows good performance; five bests and five second bests out of total ten combinations of initial stock price pairs ($S_1 (0)$, $S_2(0)$);
but, the performance of SDBS is degraded for the estimation of $\Gamma$ and $\Theta$.

From these experiments, we find that 
the differential neural network, SDBS, accurately estimates the value, the first and second derivatives
of solution function.
% weak point
But, the SDBS consumes large memory for storing many simulated paths and model parameters of NNs;
our simulation configuration needs 8GB memory of NVIDIA GTX 1080. 
The SDBS takes about 4.5 hours for 10,000 training nEpoch.
Thus, the current SDBS cannot be a practical substitute of MC simulations for the pricing of 
multidimensional European options.

\subsection{Estimation of price and Greeks of basket call options }

We consider a call option on a basket with four independent stocks.
The parameters are as in \cite{KORN} ; $T=0.5, r=0.06$, $(S_1, S_2, S_3, S_4)$ $= (40, 50, 60, 70)$, $(w_1, w_2, w_3, w_4)$ $= (0.25, 0.25, 0.25, 0.25)$, and various volatilities.
Since there is no exact solution for the basket option, 
we need a reliable numerical method to give us the benchmark value.
In Table \ref{tab:ba}, we used MC simulation based benchmark values in \cite{KORN} where the number of MC simulations paths is $10^6$.
We compare our method, SDBS used in Table ~\ref{tab:exSDBSs1s2}, with LN (the log-normal approximation of \cite{LEVY}), RG (the reciprocal gamma approximation of \cite{Milevsky}),
SLN (the shifted log-normal approximation of \cite{KORN}),
JU (the Taylor expansion approximation of \cite{Ju2002}),
and MC2 which is the MC method with $10^8$ simulated paths used in Table ~\ref{tab:exRes},
We rounded up to four decimals to be consistent with previous results in \cite{KORN}.

\begin{table}[]
\centering
\begin{tabular}{|l|c|c|c|c|c|c|c|}
\hline
K  & MC & LN & RG & SLN & JU & MC2 & SDBS \\ \hline

\multicolumn{8}{c}{$(\sigma_1, \sigma_2, \sigma_3, \sigma_4)=(0.2, 0.2, 0.2, 0.2)$} \\ \hline
\multirow{2}*{50} & 6.5355  & 6.5412 & 6.5340 & 6.5653 & 6.5404 	& 6.5407 & 6.5404 \\
  &  0 & 8.72e-04	& \textbf{2.30e-04}	& 4.56e-03	& 7.50e-04	& 7.96e-04	 & 7.50e-04 \\ \hline
  \multirow{2}*{55} & 2.5063 & 2.5104 & 2.5010 &	2.5343 &	2.5092 &	2.5094 &	2.5092 \\
  &  0 & 1.64e-03&	2.11e-03&	1.12e-02&	\textbf{1.16e-03} &	1.24e-03&	\textbf{1.16e-03} \\ \hline
  \multirow{2}*{60} & 0.5041 & 0.5037 &	0.5133 &	0.4719 &	0.5049 &	0.5049 &	0.5049\\
  &  0 & \textbf{7.93e-04} &	1.83e-02&	6.39e-02&	1.59e-03&	1.59e-03&	1.59e-03  \\ \hline
  
\multicolumn{8}{c}{$(\sigma_1, \sigma_2, \sigma_3, \sigma_4)=(0.5, 0.5, 0.5, 0.5)$} \\ \hline
\multirow{2}*{55} & 4.8324 &4.8499 &	4.7920& 	4.9492& 	4.8384& 	4.8382& 	4.8377\\
  &  0 & 3.62e-03&	8.36e-03&	2.42e-02&	1.24e-03&	1.20e-03&	\textbf{1.10e-03}  \\ \hline
  \multirow{2}*{60} & 2.7402 & 2.7463 &	2.7444 &	2.6729 &	2.7450 &	2.7441& 	2.7436 \\
  &  0 & 2.23e-03	&1.53e-03	&2.46e-02	&1.75e-03	&1.42e-03	& \textbf{1.25e-03} \\ \hline
  \multirow{2}*{65} & 1.4468 & 1.4413 &	1.4831 &	1.2550 &	1.4488 &	1.4479 &	1.4476 \\
  &  0 &3.80e-03&	2.51e-02	&1.33e-01&	1.38e-03&	7.60e-04&	\textbf{5.48e-04} \\ \hline
  
\multicolumn{8}{c}{$(\sigma_1, \sigma_2, \sigma_3, \sigma_4)=(0.8, 0.8, 0.8, 0.8)$} \\ \hline
\multirow{2}*{60} & 5.3401 & 5.3897 &5.2725 &	5.3819 &	5.3563 &	5.3468 &	5.3457 \\
 &  0 &9.29e-03&	1.27e-02&	7.83e-03&	3.03e-03	&1.25e-03&	\textbf{1.05e-03} \\ \hline
  \multirow{2}*{65} & 3.8179 & 3.8418 &	3.8123& 	3.5776& 	3.8336& 	3.8230 &	3.8215 \\
  &  0 & 6.26e-03	&1.47e-03	&6.29e-02	&4.11e-03	&1.34e-03&	\textbf{9.45e-04} \\ \hline
  \multirow{2}*{70} & 2.7011 &2.7003 &	2.7430 &	2.2590 &	2.7135& 	2.7025& 	2.7019 \\
  &  0 & \textbf{2.96e-04} &	1.55e-02&	1.64e-01&	4.59e-03&	5.18e-04&	2.98e-04 \\ \hline
  
\multicolumn{8}{c}{$(\sigma_1, \sigma_2, \sigma_3, \sigma_4)=(0.6, 1.2, 0.3, 0.9)$} \\ \hline
\multirow{2}*{60} & 5.5569 & 5.9128 &	5.7558& 	5.9371 &	5.5922& 	5.5635& 	5.5619 \\
  &  0 & 6.40e-02&	3.58e-02&	6.84e-02&	6.35e-03&	1.19e-03&	\textbf{8.98e-04} \\ \hline
  \multirow{2}*{65} & 4.1555 & 4.3459 &	4.2836 &	4.0874 &	4.1973 &	4.1604 &	4.1588 \\
  &  0 & 4.58e-02	&3.08e-02	&1.64e-02	&1.01e-02	&1.18e-03&	\textbf{8.03e-04} \\ \hline
  \multirow{2}*{70} & 3.1196 &3.1607 &	3.1798 &	2.6941 &	3.1710 &	3.1222 &	3.1207 \\
  &  0 & 1.32e-02	&1.93e-02	&1.36e-01	&1.65e-02	&8.33e-04	& \textbf{3.62e-04} \\ \hline
  
\end{tabular}
\caption{Basket call option prices and rErrors (the upper and lower values in each cell). The best results are highlighted in bold face.}
\label{tab:ba}
\end{table}

Table \ref{tab:ba} shows that SDBS gives the best accurate estimation.
Out of 12 combinations of $K$ and $(\sigma_1, \sigma_2, \sigma_3, \sigma_4)$,
SDBS gives 8 bests and 1 equal best, LN 2 bests, RG 1 best, and JU 1 equal best performances.
SDBS gives good performance for all levels of strike prices;
it gives more accurate price of options with increasing volatilities and
non-uniform volatilities.
The simulation based methods such as MC2 and SDBS can effectively process the options with large and non-uniform volatilities accurately, while the analytic approximation methods like LN, RG, SLN, and JU are degraded with such volatilities.
This trend is consistent with the results in \cite{KORN}.

\begin{table}[]
\centering
\small
\begin{tabular}{|c|c|c|c|c|c|c|c|}
\hline
  \multicolumn{2}{|c|}{ $(\sigma_1, \sigma_2, \sigma_3, \sigma_4)$} & \multicolumn{3}{|c|}{ (0.5, 0.5, 0.5, 0.5) }  &  \multicolumn{3}{|c|}{ (0.6, 1.2, 0.3, 0.9)}\\ \hline
  \multicolumn{2}{|c|}{$K$} & 55 &  60 & 65 & 60 & 65 & 70 \\
\hline
\multirow{3}*{$\Delta_1$ } &JU  &2.15e-04 & \textbf{1.17e-04} & 2.33e-04&5.73e-03 &5.63e-03 & 4.44e-03\\
& MC2  & \textbf{4.01e-05} &1.26e-04 & \textbf{1.08e-04}& \textbf{1.58e-04} & \textbf{8.17e-06} &1.69e-04\\
& SDBS  & 5.31e-04& 2.86e-04& 4.49e-04&5.19e-04   & 7.49e-05& \textbf{7.32e-05} \\ \hline

\multirow{3}*{$\Delta_2$ } &JU  &9.75e-05 & \textbf{1.88e-05} &2.13e-04 &8.27e-03 &7.92e-03 &  3.32e-03\\
& MC2  &6.57e-05 &1.51e-04 &2.11e-04 & 1.36e-04& 9.60e-05&1.98e-04\\
& SDBS  &\textbf{4.51e-05} & 1.29e-04 & \textbf{8.88e-05} & \textbf{1.19e-04}  & \textbf{2.03e-05} & \textbf{6.44e-05}\\ \hline

\multirow{3}*{$\Delta_3$ } &JU  & 1.28e-04& \textbf{2.31e-05}&4.09e-04 &2.52e-03 & 8.49e-04&3.05e-03\\
& MC2  & \textbf{1.99e-06}&9.24e-05 & \textbf{1.08e-06} & \textbf{9.81e-05} & \textbf{5.51e-05} &\textbf{1.27e-04}\\
& SDBS  & 5.57e-05&9.23e-05 & 2.46e-04& 5.15e-04&  6.89e-04& 7.48e-04\\ \hline

\multirow{3}*{$\Delta_4$ } &JU  &5.34e-04 &2.27e-04 & 7.55e-04&1.09e-03 &9.20e-03 &1.87e-02 \\
& MC2  & \textbf{8.03e-05} &2.37e-04 & 2.93e-04& 2.49e-04& \textbf{2.17e-04}& \textbf{3.11e-04} \\
& SDBS  &1.10e-04 & \textbf{2.42e-05} & \textbf{1.73e-04} & \textbf{6.00e-05} & \textbf{2.17e-04}&3.85e-04\\ \hline

\multirow{3}*{$\Gamma_1$} & JU   &2.47e-03 & 1.84e-03& 1.49e-03& 5.22e-03&8.22e-03 &1.43e-02\\
& MC2   & \textbf{3.34e-04}&\textbf{1.19e-04} &\textbf{1.12e-04} & \textbf{2.89e-05}&\textbf{1.13e-03} &\textbf{7.64e-04}\\
& SDBS  &7.99e-03 &2.98e-03 &7.61e-03 & 2.29e-03 &9.91e-03 &6.70e-03 \\ \hline

\multirow{3}*{$\Gamma_2$} & JU   &2.96e-03 & 3.81e-03& 4.10e-03&1.38e-01 &1.50e-01 &1.54e-01\\
& MC2   &2.01e-04 &\textbf{2.32e-04} &\textbf{1.25e-04} & \textbf{3.43e-04}&3.58e-04 & \textbf{1.96e-04}\\
& SDBS  & \textbf{9.01e-05}& 1.38e-03& 1.11e-03&2.20e-03 & \textbf{2.95e-04}&3.91e-04 \\ \hline

\multirow{3}*{$\Gamma_3$} & JU   & 1.25e-04& 8.40e-04& 1.70e-03&1.96e-02 & 2.24e-02&1.28e-02\\ 
& MC2   & \textbf{1.86e-05}& \textbf{5.42e-05}& \textbf{4.13e-04}& \textbf{1.01e-04}& \textbf{9.00e-04}&\textbf{5.93e-04}\\
& SDBS  & 1.23e-03& 8.23e-04&2.08e-03 &8.70e-04 & 1.03e-02&6.38e-03 \\ \hline

\multirow{3}*{$\Gamma_4$} & JU   & 1.16e-03& \textbf{1.01e-04}&8.68e-04 & 6.16e-03& 3.64e-03&6.54e-03\\
& MC2   & \textbf{2.02e-04}& 5.04e-04&\textbf{6.97e-04} &\textbf{1.94e-04} &\textbf{2.16e-04} &\textbf{2.89e-04}\\
& SDBS  & 3.81e-04& 9.63e-04& 2.49e-03&8.83e-04 & 8.10e-04&1.89e-03 \\ \hline

\multirow{3}*{$\Theta$} & JU  & 1.44e+00&8.76e-01 &5.12e-01 &1.40e-01 & 5.32e-02&1.91e-01\\
& MC2  &\textbf{1.35e-04} &\textbf{1.79e-04} & \textbf{1.85e-04}&\textbf{3.20e-04} & \textbf{3.26e-04}&3.71e-04\\
& SDBS  &1.59e-03 &1.95e-04 &1.95e-03 & 4.16e-04& 1.14e-03&\textbf{4.98e-05} \\ \hline
\end{tabular}
\caption{rErrors of Greeks estimation of basket options. The best results are highlighted in bold face. }
\label{tab:baGreeks}
\end{table}

Table \ref{tab:baGreeks} shows the rErrors of Greeks estimation obtained with the methods JU, MC2 and SDBS used in Table ~\ref{tab:ba}.
Since benchmark values are required for the calculation of the relative errors of Greeks estimations, large number  of MC simulated paths(=$10^{11}$) were generated 
and Greeks were estimated with the formulas shown in Table ~\ref{tab:Greeks}.
In $\Delta$ estimation experiments, MC2 gives 11 best and 1 equal best, and SDBS 10 best and 1 equal best, JU 2 best  performances in 24 combinations of  $K$ and $(\sigma_1, \sigma_2, \sigma_3, \sigma_4)$.
In estimation experiments of $\Gamma$ and $\Theta$, 
MC2 gives 26 best, SDBS 3 best, and  JU 1 best performances in 30 combinations of  $K$ and $(\sigma_1, \sigma_2, \sigma_3, \sigma_4)$.
The estimation performance of SDBS is high for $\Delta$, but low for $\Gamma$ and $\Theta$;
this trend is consistent with the results on exchange options in Table ~\ref{tab:exSDBSs1s2}.

As can be seen in Table ~\ref{tab:ba} and \ref{tab:baGreeks}, 
the SDBS estimates accurately for price and $\Delta$ of basket option with four assets;
for the estimation of $\Gamma$ and $\Theta$, the SDBS is worse than MC simulation methods 
but better than analytic approximation methods.

\subsection{Effect of the loss function of the Black-Scholes equation}

We expect that the loss $L_{BS}$ in Eq.~\eqref{eq:lossD} constrains the SDBS to satisfy the Black-Scholes partial differential equation; the derivatives of SDBS, Greeks, are estimated more accurately with $L_{BS}$ than without it. 
To validate the effectiveness of this constraint,
we test several weights ($w=10^{-3}, 10^{-2}$, $10^{-1}, 1, 10^1$, $10^2, 10^3$) in weighted loss $L_w$
\begin{equation}
L_w = L_{SDE} + w * L_{BS} + L_T\label{eq:Lw},
\end{equation}
where $w = 1$ corresponds to the loss in Eq.~\eqref{eq:loss};
small $w$ means small contribution of PDE constraint and relatively large contribution of the SDE,
and large $w$ means large contribution of PDE constraint and relatively small contribution of the SDE.
We perform the same experiments as in Table \ref{tab:exSDBSs1s2} with $(S_1, S_2 )=(20, 60), (60,60)$, and $(60,20)$ 
except that this experiment uses the weighted loss $L_w$ instead of the original loss in Eq.~\eqref{eq:loss}.

\begin{table}[t]
\centering
\small
\begin{tabular}{|c|c|c|c|c|c|c|c|} 
\hline
$w$  & $10^{-3}$ &  $10^{-2}$ &  $10^{-1}$ & 1 &  $10^{1}$ &  $10^2 $ &  $10^3 $ \\
\hline

\multicolumn{8}{c}{$(S_1, S_2 )=(20, 60)$} \\ \hline
Price          &4.49e-02  &3.35e-02 &  1.57e-02 &  \textbf{1.25e-02} & 2.67e-02  &5.55e-02 &1.81e-01  \\ \hline
$\Delta$     &1.92e-02  &7.05e-03 &5.91e-03 & \textbf{2.22e-03} & 1.45e-02 &1.07e-01 & 2.96e-01  \\ \hline
$\Gamma$ & 5.65e-02 & 1.12e-02& 5.22e-03 &  \textbf{1.36e-03} & 7.56e-03  &4.02e-01&6.03e-01   \\ \hline
$\Theta$    &5.82e-02  &2.78e-02 & 3.70e-03 &   \textbf{8.80e-04} & 9.64e-03  & 3.73e-01&5.74e-01 \\ \hline

\multicolumn{8}{c}{$(S_1, S_2 )=(60, 60)$} \\ \hline
Price   & 2.84e-05 &5.56e-05 & \textbf{2.97e-06} & 7.85e-06 & 1.19e-05 &3.97e-04 &2.50e-04 \\ \hline
$\Delta$  &5.29e-04 & 3.90e-05  &4.99e-05 & \textbf{9.44e-06}  &4.56e-05 &5.46e-04 &2.11e-03\\ \hline
$\Gamma$ &1.25e-01 &1.61e-02  &3.55e-03 & 1.52e-03 & \textbf{1.12e-03}  &2.90e-03 &1.02e-02\\ \hline
$\Theta$   & 1.05e-01 &1.63e-02  &5.26e-03 &  \textbf{2.03e-03}  & 2.73e-03&4.72e-03 &4.42e-03 \\ \hline

\multicolumn{8}{c}{$(S_1, S_2 )=(60, 20)$} \\ \hline
Price          & 9.28e-06 &\textbf{4.15e-08} &  1.57e-07 & 2.90e-07&7.17e-07   &3.32e-06 & 8.15e-06 \\ \hline
$\Delta$     &1.14e-03  &1.13e-04 &7.86e-06  &  \textbf{2.25e-06}&  3.29e-06 &2.95e-05 & 1.06e-04 \\ \hline
$\Gamma$ &  6.86e+00&1.03e+00 & 2.15e-01 & 8.42e-02 & 5.11e-02  & \textbf{9.77e-03} &  1.31e-01\\ \hline
$\Theta$    &3.66e+00 & 9.36e-01& 2.66e-01 & 1.09e-01 &  5.81e-02 &  \textbf{1.83e-02} & 1.65e-01 \\ \hline

\end{tabular}
\caption{rErrors of weighted loss $L_w$ for the exchange option. The best results are highlighted in bold face.}
\label{tab:LD}
\end{table}

Table~\ref{tab:LD} shows that this version of SDBS performs accurately for wide range of $w$, but it degrades when the $w$ deviates from unity.
This means that suitable constraint of Black-Scholes equation enhances the performance of option pricing.
Price estimation performs better with $w \leq 1$ 
while $\Gamma$ and $\Theta$ estimation perform better with $w \geq 1$.
This is because weighting more to the loss of differential equation $L_{BS}$ gives accurate estimation for derivatives
while weighting more to the loss of value $L_{SDE}$ gives accurate estimation for values.
When $L_{BS}$ is removed completely, no convergent value is obtained. 
These observations reveal that the loss $L_{BS}$ is essential for the accurate estimation of gradient and Hessian of NN, and both loss $L_{BS}$ and $L_{SDE}$ play a key role in making the SDBS suitable for option pricing problems.

\subsection{Effect of the smoothness of activation functions}\label{sec:activation}

In section \ref{sec:mathActi}, we described the mathematical conditions of activation functions for the approximation of  the functions with all partial derivatives;
the boundedness and continuity of all its partial derivatives up to order $m$ are required to approximate the functions with all partial derivatives up to $m$ are continuous and bounded \cite{HORNIK1990, GALLANT1992129}.
The activation functions such as sigmoid, tanh, and sin satisfy these conditions.
But, the practical performance can differ from one activation function to another.
We compare the performance of several activation functions in terms of 
their suitability for option pricing problems.

\begin{table}[t]
\centering
\small
\begin{tabular}{|c|c|c|c|c|c|c|c|} 
\hline
  &  sigmoid & tanh & sin & relu & elu & selu & softplus \\
\hline

\multicolumn{8}{c}{$(S_1, S_2 )=(20, 60)$} \\ \hline
Price          & 8.78e-01 & 1.76e-02 &  3.93e-01  &  1.01e+00 &  1.39e-02 &  5.55e-01 &  \textbf{1.25e-02} \\ \hline
$\Delta$     & 5.06e-01 &  5.29e-03  &  9.93e-01  &  9.94e-01 &  2.01e-01 &  7.59e-01 &  \textbf{2.22e-03}\\ \hline
$\Gamma$ & 5.83e-01 &  1.90e-02  &  9.92e-01  &  1.00e+00 &  1.39e-01 &  9.51e-01 &  \textbf{1.36e-03}\\ \hline
$\Theta$    & 5.55e-01 &  3.40e-02  &  1.00e+00  &  1.00e+00 &  2.59e-01 &  9.07e-01 &  \textbf{8.80e-04}\\ \hline

\multicolumn{8}{c}{$(S_1, S_2 )=(60, 60)$} \\ \hline
Price   & \textbf{2.83e-06} & 1.43e-04 & 3.79e-04  & 8.69e-01 & 6.593-05 &1.88e-01 & 7.85e-06\\ \hline
$\Delta$  & 4.57e-05 & 3.07e-04 &  1.14e-03 & 1.02e-01 &2.973-04 & 1.42e-01 & \textbf{9.44e-06}\\ \hline
$\Gamma$ & 2.99e-03 & 3.65e-03 & 5.49e-03 & 1.00e+00 &1.813-03& 5.04e-03 & \textbf{1.52e-03}\\ \hline
$\Theta$   & 8.14e-03 & \textbf{1.90e-03} &  5.84e-03 & 9.96e-01 & 3.763-03 & 2.74e-02 & 2.03e-03\\ \hline

\multicolumn{8}{c}{$(S_1, S_2 )=(60, 20)$} \\ \hline
Price          & 6.10e-06 & 5.09e-05 &  6.08e-05  &  1.48e-04 &  1.66e-06 &  3.75e-03 &  \textbf{2.90e-07} \\ \hline
$\Delta$     & 8.50e-05 &  3.37e-04  & 4.49e-04  &  6.84e-05&  1.68e-04 &  5.05e-03 &  \textbf{2.25e-06}\\ \hline
$\Gamma$ & \textbf{8.41e-02} &  4.82e-01  &  6.48e-01  &  1.00e+00 &  2.63e-01 &  3.75e+00 & 8.42 e-02\\ \hline
$\Theta$    & 1.57e-01 &  5.39e-01  &  7.42e-01  &  1.06e+00 &  4.46e-01 &  4.04e+00 &  \textbf{1.09e-01}\\ \hline

\end{tabular}
\caption{rErrors of activation functions for exchange option. The best results are highlighted in bold face.}
\label{tab:actEx}
\end{table}

Using the same experimental setting used for SDBS in Table~\ref{tab:exSDBSs1s2} with nEpoch 100,000,
we examine the rErrors of estimation with several activation functions.
As can be seen in Table~\ref{tab:actEx}, the accuracy differs substantially from activation function to  activation function.
Sigmoid and softplus show significantly better accuracy than other methods for estimating price and $\Delta$;
but, sigmoid degraded for $(S_1, S_2 )=(20, 60)$.
The relu-like function such as relu, elu, selu, softplus
have given good performance in most applications of NNs.
But, the insufficient smoothness of relu, elu, and selu
degrade the estimation accuracy of the $\Gamma$ and $\Theta$.
These accuracies are exactly proportional to the smoothness of the relu-like functions.
Non-differentiable $C^0$ functions such as relu and selu give the lowest performance,
especially large errors( $\sim$ e-01 and e+00).
Another relu-like $C^1$ function elu gives intermediate performance,
while relu-like infinitely differentiable function softplus shows high performance. 
This is because the smoothness of the neural network is determined by the smoothness of the component functions,
and sufficient smoothness of activation functions is an essential factor for training the SDBS which is strongly based on differentiability of NNs.
The accuracy of price estimation by relu-like functions in Eq.~\eqref{eq:reluAct}  is comparable to
that by $C^\infty$ functions in Eq.~\eqref{eq:smoothAct};
for example, elu shows good performance for estimating price comparing with $C^\infty$ functions.
This is because the vanishing gradient problem is reduced for relu-like functions.

The above considerations suggest that relu-like functions are required for 
the estimation of accurate price
and $C^\infty$ functions are required for accurate Greeks estimation.
Hence, the intersection of relu-like functions in Eq.~\eqref{eq:reluAct} and  $C^\infty$ functions in Eq.~\eqref{eq:smoothAct}, softplus, gives the best performance for the present work.

\section{Conclusions}
% summary, strong point
We improve the approximation capability of NNs by making NNs have the differential relationship of the problem we wish to solve. Following this approach, the proposed method learns the Black-Scholes equation of option price.
The sufficient smoothness of activation functions is an essential factor 
for the proposed method which relies heavily on the differentiability of NNs;
The softplus activation function is suitable for our method.

The SDBS is trained using more realistic asset price paths generated by SDEs.
Hence, the SDBS can accurately model option price  and
does not suffer from the curse of dimensionality associated with high dimensional FDMs.
Since the SDBS utilizes the exact differential relationship along a single simulation path,
it can easily use parallel and distributed computing.
We can also use this single-path dependency for backward recursion in American option pricing which will be discussed in a separate paper.

% future work
Further improvements can be made using sample paths
generated by advanced MC simulation methods such as variance reduction techniques and low-discrepancy sequences.
Although the proposed method was applied to a typical Black-Scholes model of option pricing in this work, it can be modified to suit different models and option types.
In addition, differential neural networks can be applied to solve problems related to partial differential equations such as inverse problems, the Navier-Stokes equation, and the Schr{\"o}dinger equation.

\section*{Appendix}

In this section we derive the calculation formulas of Greeks in Table~\ref{tab:Greeks}.
Although the exchange option has the exact solution, we prepare the MC approximation of Greeks for comparing with other numerical methods. Pathwise derivative estimate (PW) method first differentiates  $e^{-r(T-t)}V(T) $ in Eq.\eqref{eq:mcPrice}
and then expectation is performed \cite{glasserman}, i.e., differentiation and expectation is interchanged.
For the delta of exchange option, $e^{-r(T-t)} [ S_1(T) - S_2(T) ]\mathbbm{1} \{S_1(T) > S_2(T) \}$ at $t=0$ is differentiated with respect to $S_1(0)$,
\begin{align}
\frac{d}{d S_1(0)} e^{-rT} V(T)   &= e^{-rT}  \frac{dS_1(T)}{d S_1(0)} \frac{dV(T)}{d S_1(T)}   \nonumber \\
&=  e^{-rT} \frac{S_1(T)}{S_1(0)}  \mathbbm{1} \{S_1(T) > S_2(T) \} ,\label{eq:deltaEx}
\end{align}
where $S_1 (T)$ is presented in Eq.~\eqref{eq:stock} and  $\mathbbm{1}$ denotes a indicator function.

Similarly to the calculation of delta, the theta of  exchange option is obtained by the expectation of the following,
\begin{align}
& \frac{\partial e^{-r(T-t)} V(T-t)}{\partial t} |_{t=0} =
    \frac{\partial e^{-r(T-t)} }{\partial t} |_{t=0} V(T) + e^{-rT}\frac{\partial V(T-t)}{\partial t} |_{t=0} \nonumber \\
& = r  e^{-r T} V(T) +   e^{-r T}  \frac{\partial [ S_1 (T-t) - S_2 (T-t) ] \mathbbm{1} \{S_1(T-t) > S_2(T-t) \}  }{\partial t} |_{t=0} \nonumber \\
& = e^{-rT} \big( r [S_1(T) - S_2(T) ] \nonumber \\
& \quad - [ S_1(T) ( r - \frac{\sigma_1^2}{2} + \frac{\sigma_1 L_1 Z}{2 \sqrt{T} } ) - S_2(T) ( r - \frac{\sigma_2^2}{2} + \frac{\sigma_2 L_2 Z}{2 \sqrt{T} } ) ]  \big) \mathbbm{1} \{S_1(T) > S_2(T) \} \nonumber \\
& = e^{-rT} [ S_1(T) ( \frac{\sigma_1^2}{2} - \frac{\sigma_1 L_1 Z}{2 \sqrt{T} } ) 
- S_2(T) ( \frac{\sigma_2^2}{2} - \frac{\sigma_2 L_2 Z}{2 \sqrt{T} } ) ] \mathbbm{1} \{S_1(T) > S_2(T) \}
\label{eq:thetaEx}
\end{align}
where $L_i$ and $Z$ are described in ~\eqref{eq:stock}.

We also use the PW method for pricing basket call option.  
The delta of basket option is calculated with $V(T)$ in Eq.\eqref{eq:BasCall}.
The delta is given by the expectation of
\begin{equation}
\frac{d e^{-rT} V(T)}{d S_i (0)} = e^{-rT} \frac{S_i (T)}{S_i (0)}  w_i \mathbbm{1} \{ \sum_{i=1}^{n} w_i S_i(T) > K \}
\end{equation} 

The theta of  basket call option is given by the expectation of
\begin{align}
& \frac{\partial e^{-r(T-t)} V(T-t)}{\partial t} |_{t=0} = \nonumber \\
& e^{-rT} [  -rK + \sum_{i=1}^{n} w_i S_i(T) ( \frac{\sigma_i^2}{2} - \frac{\sigma_i L_i Z}{2 \sqrt{T} } ) ]
 \mathbbm{1} \{ \sum_{i=1}^{n} w_i S_i(T) > K \}
\end{align}

%LR
For the calculation of the gamma using PW method,  the first derivatives of payoff functions are required to be differentiable functions of the parameter $S_i(0)$.
But, the first derivatives, the delta, are not differentiable for exchange and basket options and
PW cannot be applied for these options.
Likelihood ratio method (LR), an alternate method, assumes 
that the assets $S_i$ has a probability density $g_\alpha$ 
and that $\alpha$ denotes a parameter of this density.
To emphasize that the expectation is computed with respect to $g_\alpha$,
$E_\alpha$ is used.
LR interchange the order of differentiation and integration to derive a Greeks
\begin{align}
& \frac{d}{d\alpha}  E_\alpha [V(T)] = \int_{R^n} V(T) \frac{d}{d\alpha} g_\alpha (x) dx  \nonumber \\
& = \int_{R^n} V(T) \frac{\dot g_\alpha(x)}{g_\alpha(x)} g_\alpha (x) dx = E_\alpha [V(T)\frac{\dot g_\alpha(x)}{g_\alpha(x)}]
\end{align}

In the present work, Eq.\eqref{eq:stock} suggests that  $S_i(T)$ for $i=1,\ldots,n$, has the distribution $\exp(Y_i)$.
The vector $Y=(Y_1, Y_2, \ldots, Y_n)^\prime $ follows the normal distribution 
$N(\mu (\alpha), \tilde \Sigma (\alpha))$,
where $\mu (\alpha;\cdot)$ denotes a vector with elements 
$\mu_i = \ln(S_i(0)) +(r-\frac{1}{2} \sigma_i^2)T$,
$\tilde \Sigma (\alpha) = T A\Sigma A^\prime$
($A=diag(\sigma_1, \ldots, \sigma_n)$), and $\alpha = (S_1(0), \ldots, S_n(0))$.
The probability density function of $Y$
could be written by
\begin{equation}
g_\alpha (Y) = \frac{1}{\sqrt{(2\pi)^n | \tilde \Sigma (\alpha) | }}  
\exp [ - \frac{1}{2 } (y-\mu (\alpha))^{\prime} \tilde \Sigma (\alpha)^{-1} (y-\mu (\alpha))].
\label{eq:multiN}
\end{equation} 
We generate the sample paths $Y$ by 
$\mu + \sqrt{T}ALZ$, where $LL^\prime=\Sigma$, and $Z$ denotes a standard normal distribution.
Hence,
\begin{align}
\frac{\dot g_{S_i(0)}}{g_{S_i(0)}} &= \frac{d}{d S_i(0)} \log(g_{S_i(0)}(Y) )
= (Y-\mu)^{\prime} \tilde \Sigma^{-1} \frac{d\mu}{dS_i(0)}  
= (\sqrt{T} A L Z)^{\prime}   \tilde\Sigma^{-1}  \frac{d\mu}{dS_i(0)} \nonumber \\
&= (\sqrt{T} A L Z)^{\prime}   (T A\Sigma A^\prime)^{-1}  (0, \cdots, \frac{1}{S_i(0)}, \cdots, 0)^\prime
= \frac{ ( Z^{\prime} L^{-1} A^{-1} )_i }{\sqrt{T} S_i(0)}
\end{align}
where $ ( Z^{\prime} L^{-1} A^{-1} )_i$ denotes the $i$th component of the row vector
 $Z^{\prime} L^{-1} A^{-1}$.

For calculating gamma in the present study, LR method is first applied to the delta calculation
and this delta is differentiated and averaged by the PW method.
This LR-PW method for gamma calculation gives superior performance
compared with the pure LR method in \cite{glasserman}.

For the exchange option, the gamma is given by the expectation of the following expression
\begin{align}
\frac{d }{dS_1(0) }
& \big( e^{-rT} \max [ S_1(T) - S_2(T), 0 ] \frac{ ( Z^{\prime} L^{-1} A^{-1} )_1 }{\sqrt{T} S_1(0)} \big) \nonumber \\
&=e^{-rT} \big( \frac{d \max [ S_1(T) - S_2(T), 0 ] }{dS_1(0) } \big) \frac{ ( Z^{\prime} L^{-1} A^{-1} )_1 }{\sqrt{T} S_1(0)} 
\nonumber \\
& \quad + e^{-rT} \max [ S_1(T) - S_2(T), 0 ] \frac{d  }{dS_1(0) } \big( \frac{ ( Z^{\prime} L^{-1} A^{-1} )_1 }{\sqrt{T} S_1(0)}  \big) \nonumber \\
&=e^{-rT}  \mathbbm{1} \{(S_1(T) > S_2(T) \frac{ ( Z^{\prime} L^{-1} A^{-1} )_1 }{\sqrt{T} S_1(0)} 
\nonumber \\
& \quad + e^{-rT} \max [ S_1(T) - S_2(T), 0 ]   \frac{ ( Z^{\prime} L^{-1} A^{-1} )_1 }{-\sqrt{T} S_1^2(0)}  \nonumber \\
&=  e^{-rT}  \frac{ ( Z^{\prime} L^{-1} A^{-1} )_1 }{\sqrt{T} }  \frac{ S_2 (T) }{ S_1^2(0)}
 \mathbbm{1} \{(S_1(T) > S_2(T) \}.\label{eq:gammaEx}
\end{align}

For the basket call option, the gamma is given by the expectation of the following expression
\begin{align}
&\frac{d }{dS_i(0) } [ e^{-rT} \max( \sum_{i=1}^{n} w_i S_i(T) - K,0)
\frac{ ( Z^{\prime} L^{-1} A^{-1} )_i }{\sqrt{T} S_i(0)} ] \nonumber \\
&= e^{-rT}  \frac{d }{dS_i(0) } [ \max( \sum_{i=1}^{n} w_i S_i(T) - K,0) ]
\frac{ ( Z^{\prime} L^{-1} A^{-1} )_i }{\sqrt{T} S_i(0)} \nonumber \\
&\quad +  e^{-rT} \max( \sum_{i=1}^{n} w_i S_i(T) - K,0)
\frac{d }{dS_i(0) } [ \frac{ ( Z^{\prime} L^{-1} A^{-1} )_i }{\sqrt{T} S_i(0)} ] \nonumber \\
&= e^{-rT} w_i  \mathbbm{1} \{ \sum_{i=1}^{n} w_i S_i(T) > K  \} 
\frac{ ( Z^{\prime} L^{-1} A^{-1} )_i }{\sqrt{T} S_i(0)} \nonumber \\
&\quad +  e^{-rT} \max( \sum_{i=1}^{n} w_i S_i(T) - K,0)
 [ \frac{ ( Z^{\prime} L^{-1} A^{-1} )_i }{ -\sqrt{T} S_i^2(0)} ] \nonumber \\
&=  e^{-rT}  \frac{ ( Z^{\prime} L^{-1} A^{-1} )_i }{\sqrt{T} S_i^2(0) }   
\{ w_i S_i(T) - \sum_{i=1}^{n} w_i S_i(T) + K \}
 \mathbbm{1} \{ \sum_{i=1}^{n} w_i S_i(T) > K \}.\label{eq:gammaBa}
\end{align}

\bibliography{mybibfile}

\begin{thebibliography}{31}
\expandafter\ifx\csname natexlab\endcsname\relax\def\natexlab#1{#1}\fi
\providecommand{\url}[1]{\texttt{#1}}
\providecommand{\href}[2]{#2}
\providecommand{\path}[1]{#1}
\providecommand{\DOIprefix}{doi:}
\providecommand{\ArXivprefix}{arXiv:}
\providecommand{\URLprefix}{URL: }
\providecommand{\Pubmedprefix}{pmid:}
\providecommand{\doi}[1]{\href{http://dx.doi.org/#1}{\path{#1}}}
\providecommand{\Pubmed}[1]{\href{pmid:#1}{\path{#1}}}
\providecommand{\bibinfo}[2]{#2}
\ifx\xfnm\relax \def\xfnm[#1]{\unskip,\space#1}\fi
%Type = Misc
\bibitem[{Abadi et~al.(2015)Abadi, Agarwal, Barham, Brevdo, Chen, Citro,
  Corrado, Davis, Dean, Devin, Ghemawat, Goodfellow, Harp, Irving, Isard, Jia,
  Jozefowicz, Kaiser, Kudlur, Levenberg, Man\'{e}, Monga, Moore, Murray, Olah,
  Schuster, Shlens, Steiner, Sutskever, Talwar, Tucker, Vanhoucke, Vasudevan,
  Vi\'{e}gas, Vinyals, Warden, Wattenberg, Wicke, Yu and Zheng}]{tf}
\bibinfo{author}{Abadi, M.}, \bibinfo{author}{Agarwal, A.},
  \bibinfo{author}{Barham, P.}, \bibinfo{author}{Brevdo, E.},
  \bibinfo{author}{Chen, Z.}, \bibinfo{author}{Citro, C.},
  \bibinfo{author}{Corrado, G.S.}, \bibinfo{author}{Davis, A.},
  \bibinfo{author}{Dean, J.}, \bibinfo{author}{Devin, M.},
  \bibinfo{author}{Ghemawat, S.}, \bibinfo{author}{Goodfellow, I.},
  \bibinfo{author}{Harp, A.}, \bibinfo{author}{Irving, G.},
  \bibinfo{author}{Isard, M.}, \bibinfo{author}{Jia, Y.},
  \bibinfo{author}{Jozefowicz, R.}, \bibinfo{author}{Kaiser, L.},
  \bibinfo{author}{Kudlur, M.}, \bibinfo{author}{Levenberg, J.},
  \bibinfo{author}{Man\'{e}, D.}, \bibinfo{author}{Monga, R.},
  \bibinfo{author}{Moore, S.}, \bibinfo{author}{Murray, D.},
  \bibinfo{author}{Olah, C.}, \bibinfo{author}{Schuster, M.},
  \bibinfo{author}{Shlens, J.}, \bibinfo{author}{Steiner, B.},
  \bibinfo{author}{Sutskever, I.}, \bibinfo{author}{Talwar, K.},
  \bibinfo{author}{Tucker, P.}, \bibinfo{author}{Vanhoucke, V.},
  \bibinfo{author}{Vasudevan, V.}, \bibinfo{author}{Vi\'{e}gas, F.},
  \bibinfo{author}{Vinyals, O.}, \bibinfo{author}{Warden, P.},
  \bibinfo{author}{Wattenberg, M.}, \bibinfo{author}{Wicke, M.},
  \bibinfo{author}{Yu, Y.}, \bibinfo{author}{Zheng, X.}, \bibinfo{year}{2015}.
\newblock \bibinfo{title}{{TensorFlow}: Large-scale machine learning on
  heterogeneous systems}.
\newblock \URLprefix \url{https://www.tensorflow.org/}. \bibinfo{note}{software
  available from tensorflow.org}.
%Type = Article
\bibitem[{Becker et~al.(2019)Becker, Cheridito, Jentzen and Welti}]{Becker2019}
\bibinfo{author}{Becker, S.}, \bibinfo{author}{Cheridito, P.},
  \bibinfo{author}{Jentzen, A.}, \bibinfo{author}{Welti, T.},
  \bibinfo{year}{2019}.
\newblock \bibinfo{title}{Solving high-dimensional optimal stopping problems
  using deep learning.}
\newblock \bibinfo{journal}{CoRR} \bibinfo{volume}{abs/1908.01602}.
\newblock \URLprefix
  \url{http://dblp.uni-trier.de/db/journals/corr/corr1908.html#abs-1908-01602}.
%Type = Article
\bibitem[{Black and Scholes(1973)}]{Black1973}
\bibinfo{author}{Black, F.}, \bibinfo{author}{Scholes, M.},
  \bibinfo{year}{1973}.
\newblock \bibinfo{title}{The pricing of options and corporate liabilityies}.
\newblock \bibinfo{journal}{the Journal of Political Economy}
  \bibinfo{volume}{81}, \bibinfo{pages}{637--654}.
\newblock \DOIprefix\doi{10.1086/260062}.
%Type = Article
\bibitem[{Carr et~al.(2001)Carr, Stanley and Madan}]{Carr1999}
\bibinfo{author}{Carr, P.}, \bibinfo{author}{Stanley, M.},
  \bibinfo{author}{Madan, D.}, \bibinfo{year}{2001}.
\newblock \bibinfo{title}{Option valuation using the fast fourier transform}.
\newblock \bibinfo{journal}{J. Comput. Finance} \bibinfo{volume}{2}.
\newblock \DOIprefix\doi{10.21314/JCF.1999.043}.
%Type = Article
\bibitem[{Clevert et~al.(2015)Clevert, Unterthiner and Hochreiter}]{elu}
\bibinfo{author}{Clevert, D.A.}, \bibinfo{author}{Unterthiner, T.},
  \bibinfo{author}{Hochreiter, S.}, \bibinfo{year}{2015}.
\newblock \bibinfo{title}{Fast and accurate deep network learning by
  exponential linear units (elus)}.
\newblock \bibinfo{journal}{Under Review of ICLR2016 (1997)} .
%Type = Article
\bibitem[{Cybenko(1989)}]{Cybenko}
\bibinfo{author}{Cybenko, G.}, \bibinfo{year}{1989}.
\newblock \bibinfo{title}{{Approximation by superpositions of a sigmoidal
  function}}.
\newblock \bibinfo{journal}{Mathematics of Control, Signals, and Systems
  (MCSS)} \bibinfo{volume}{2}, \bibinfo{pages}{303--314}.
\newblock \URLprefix \url{http://dx.doi.org/10.1007/BF02551274},
  \DOIprefix\doi{10.1007/BF02551274}.
%Type = Techreport
\bibitem[{Deng and Yu(2014)}]{deng}
\bibinfo{author}{Deng, L.}, \bibinfo{author}{Yu, D.}, \bibinfo{year}{2014}.
\newblock \bibinfo{title}{Deep Learning: Methods and Applications}.
\newblock \bibinfo{type}{Technical Report} \bibinfo{number}{MSR-TR-2014-21}.
\newblock \URLprefix
  \url{https://www.microsoft.com/en-us/research/publication/deep-learning-methods-and-applications/}.
%Type = Article
\bibitem[{E et~al.(2017)E, Han and Jentzen}]{Wei2017}
\bibinfo{author}{E, W.}, \bibinfo{author}{Han, J.}, \bibinfo{author}{Jentzen,
  A.}, \bibinfo{year}{2017}.
\newblock \bibinfo{title}{Deep learning-based numerical methods for
  high-dimensional parabolic partial differential equations and backward
  stochastic differential equations}.
\newblock \bibinfo{journal}{Communications in Mathematics and Statistics}
  \bibinfo{volume}{5}, \bibinfo{pages}{349--380}.
\newblock \DOIprefix\doi{10.1007/s40304-017-0117-6}.
%Type = Article
\bibitem[{Gallant and White(1992)}]{GALLANT1992129}
\bibinfo{author}{Gallant, A.R.}, \bibinfo{author}{White, H.},
  \bibinfo{year}{1992}.
\newblock \bibinfo{title}{On learning the derivatives of an unknown mapping
  with multilayer feedforward networks}.
\newblock \bibinfo{journal}{Neural Networks} \bibinfo{volume}{5},
  \bibinfo{pages}{129 -- 138}.
\newblock \URLprefix
  \url{http://www.sciencedirect.com/science/article/pii/S0893608005800115},
  \DOIprefix\doi{https://doi.org/10.1016/S0893-6080(05)80011-5}.
%Type = Article
\bibitem[{{Gencay} and {Min Qi}(2001)}]{Gencay2001}
\bibinfo{author}{{Gencay}, R.}, \bibinfo{author}{{Min Qi}},
  \bibinfo{year}{2001}.
\newblock \bibinfo{title}{Pricing and hedging derivative securities with neural
  networks: Bayesian regularization, early stopping, and bagging}.
\newblock \bibinfo{journal}{IEEE Transactions on Neural Networks}
  \bibinfo{volume}{12}, \bibinfo{pages}{726--734}.
\newblock \DOIprefix\doi{10.1109/72.935086}.
%Type = Book
\bibitem[{Glasserman(2004)}]{glasserman}
\bibinfo{author}{Glasserman, P.}, \bibinfo{year}{2004}.
\newblock \bibinfo{title}{Monte Carlo methods in financial engineering}.
\newblock \bibinfo{publisher}{Springer}, \bibinfo{address}{New York}.
\newblock \URLprefix
  \url{http://www.amazon.com/Financial-Engineering-Stochastic-Modelling-Probability/dp/0387004513/ref=pd_sim_b_68?ie=UTF8&refRID=1AN8JXSDGMEV2RPHFC2A}.
%Type = Inproceedings
\bibitem[{Glorot et~al.(2011)Glorot, Bordes and Bengio}]{softplus}
\bibinfo{author}{Glorot, X.}, \bibinfo{author}{Bordes, A.},
  \bibinfo{author}{Bengio, Y.}, \bibinfo{year}{2011}.
\newblock \bibinfo{title}{Deep sparse rectifier neural networks}, in:
  \bibinfo{editor}{Gordon, G.}, \bibinfo{editor}{Dunson, D.},
  \bibinfo{editor}{Dudík, M.} (Eds.), \bibinfo{booktitle}{Proceedings of the
  Fourteenth International Conference on Artificial Intelligence and
  Statistics}, \bibinfo{publisher}{PMLR}, \bibinfo{address}{Fort Lauderdale,
  FL, USA}. pp. \bibinfo{pages}{315--323}.
\newblock \URLprefix \url{http://proceedings.mlr.press/v15/glorot11a.html}.
%Type = Incollection
\bibitem[{Graves and Schmidhuber(2009)}]{jugen}
\bibinfo{author}{Graves, A.}, \bibinfo{author}{Schmidhuber, J.},
  \bibinfo{year}{2009}.
\newblock \bibinfo{title}{Offline handwriting recognition with multidimensional
  recurrent neural networks}, in: \bibinfo{editor}{Koller, D.},
  \bibinfo{editor}{Schuurmans, D.}, \bibinfo{editor}{Bengio, Y.},
  \bibinfo{editor}{Bottou, L.} (Eds.), \bibinfo{booktitle}{Advances in Neural
  Information Processing Systems 21}. \bibinfo{publisher}{Curran Associates,
  Inc.}, pp. \bibinfo{pages}{545--552}.
\newblock \URLprefix
  \url{http://papers.nips.cc/paper/3449-offline-handwriting-recognition-with-multidimensional-recurrent-neural-networks.pdf}.
%Type = Article
\bibitem[{Hornik et~al.(1990)Hornik, Stinchcombe and White}]{HORNIK1990}
\bibinfo{author}{Hornik, K.}, \bibinfo{author}{Stinchcombe, M.},
  \bibinfo{author}{White, H.}, \bibinfo{year}{1990}.
\newblock \bibinfo{title}{Universal approximation of an unknown mapping and its
  derivatives using multilayer feedforward networks}.
\newblock \bibinfo{journal}{Neural Networks} \bibinfo{volume}{3},
  \bibinfo{pages}{551 -- 560}.
\newblock \URLprefix
  \url{http://www.sciencedirect.com/science/article/pii/0893608090900056},
  \DOIprefix\doi{https://doi.org/10.1016/0893-6080(90)90005-6}.
%Type = Book
\bibitem[{Hull(2018)}]{Hull}
\bibinfo{author}{Hull, J.}, \bibinfo{year}{2018}.
\newblock \bibinfo{title}{Options, futures, and other derivatives}.
\newblock \bibinfo{edition}{10. ed., pearson internat. ed} ed.,
  \bibinfo{publisher}{Pearson Prentice Hall}.
%Type = Article
\bibitem[{Ju(2002)}]{Ju2002}
\bibinfo{author}{Ju, N.}, \bibinfo{year}{2002}.
\newblock \bibinfo{title}{Pricing asian and basket options via taylor
  expansion}.
\newblock \bibinfo{journal}{J. Comput. Finance} \bibinfo{volume}{5},
  \bibinfo{pages}{79--103}.
%Type = Misc
\bibitem[{Kingma and Ba(2014)}]{adam}
\bibinfo{author}{Kingma, D.P.}, \bibinfo{author}{Ba, J.}, \bibinfo{year}{2014}.
\newblock \bibinfo{title}{Adam: A method for stochastic optimization}.
\newblock \URLprefix \url{http://arxiv.org/abs/1412.6980}. \bibinfo{note}{cite
  arxiv:1412.6980Comment: Published as a conference paper at the 3rd
  International Conference for Learning Representations, San Diego, 2015}.
%Type = Incollection
\bibitem[{Klambauer et~al.(2017)Klambauer, Unterthiner, Mayr and
  Hochreiter}]{selu}
\bibinfo{author}{Klambauer, G.}, \bibinfo{author}{Unterthiner, T.},
  \bibinfo{author}{Mayr, A.}, \bibinfo{author}{Hochreiter, S.},
  \bibinfo{year}{2017}.
\newblock \bibinfo{title}{Self-normalizing neural networks}, in:
  \bibinfo{editor}{Guyon, I.}, \bibinfo{editor}{Luxburg, U.V.},
  \bibinfo{editor}{Bengio, S.}, \bibinfo{editor}{Wallach, H.},
  \bibinfo{editor}{Fergus, R.}, \bibinfo{editor}{Vishwanathan, S.},
  \bibinfo{editor}{Garnett, R.} (Eds.), \bibinfo{booktitle}{Advances in Neural
  Information Processing Systems 30}. \bibinfo{publisher}{Curran Associates,
  Inc.}, pp. \bibinfo{pages}{971--980}.
\newblock \URLprefix
  \url{http://papers.nips.cc/paper/6698-self-normalizing-neural-networks.pdf}.
%Type = Article
\bibitem[{Kohler et~al.(2010)Kohler, Krzyżak and Todorovic}]{Kohler2010}
\bibinfo{author}{Kohler, M.}, \bibinfo{author}{Krzyżak, A.},
  \bibinfo{author}{Todorovic, N.}, \bibinfo{year}{2010}.
\newblock \bibinfo{title}{Pricing of high-dimensional american options by
  neural networks}.
\newblock \bibinfo{journal}{Mathematical Finance} \bibinfo{volume}{20},
  \bibinfo{pages}{384--410}.
\newblock \DOIprefix\doi{10.1111/j.1467-9965.2010.00404.x}.
%Type = Article
\bibitem[{Korn and Zeytun(2013)}]{KORN}
\bibinfo{author}{Korn, R.}, \bibinfo{author}{Zeytun, S.}, \bibinfo{year}{2013}.
\newblock \bibinfo{title}{Efficient basket monte carlo option pricing via a
  simple analytical approximation}.
\newblock \bibinfo{journal}{Journal of Computational and Applied Mathematics}
  \bibinfo{volume}{243}, \bibinfo{pages}{48 -- 59}.
\newblock \URLprefix
  \url{http://www.sciencedirect.com/science/article/pii/S0377042712004827},
  \DOIprefix\doi{https://doi.org/10.1016/j.cam.2012.10.035}.
%Type = Inproceedings
\bibitem[{Krizhevsky et~al.(2012)Krizhevsky, Sutskever and Hinton}]{hinton}
\bibinfo{author}{Krizhevsky, A.}, \bibinfo{author}{Sutskever, I.},
  \bibinfo{author}{Hinton, G.E.}, \bibinfo{year}{2012}.
\newblock \bibinfo{title}{Imagenet classification with deep convolutional
  neural networks}, in: \bibinfo{booktitle}{Proceedings of the 25th
  International Conference on Neural Information Processing Systems - Volume
  1}, \bibinfo{publisher}{Curran Associates Inc.}, \bibinfo{address}{Red Hook,
  NY, USA}. p. \bibinfo{pages}{1097–1105}.
%Type = Inproceedings
\bibitem[{Le et~al.(2012)Le, Ranzato, Monga, Devin, Chen, Corrado, Dean and
  Ng}]{ng}
\bibinfo{author}{Le, Q.V.}, \bibinfo{author}{Ranzato, M.},
  \bibinfo{author}{Monga, R.}, \bibinfo{author}{Devin, M.},
  \bibinfo{author}{Chen, K.}, \bibinfo{author}{Corrado, G.S.},
  \bibinfo{author}{Dean, J.}, \bibinfo{author}{Ng, A.Y.}, \bibinfo{year}{2012}.
\newblock \bibinfo{title}{Building high-level features using large scale
  unsupervised learning}, in: \bibinfo{booktitle}{Proceedings of the 29th
  International Coference on International Conference on Machine Learning},
  \bibinfo{publisher}{Omnipress}, \bibinfo{address}{Madison, WI, USA}. p.
  \bibinfo{pages}{507–514}.
%Type = Article
\bibitem[{Levy(1992)}]{LEVY}
\bibinfo{author}{Levy, E.}, \bibinfo{year}{1992}.
\newblock \bibinfo{title}{Pricing european average rate currency options}.
\newblock \bibinfo{journal}{Journal of International Money and Finance}
  \bibinfo{volume}{11}, \bibinfo{pages}{474 -- 491}.
\newblock \URLprefix
  \url{http://www.sciencedirect.com/science/article/pii/026156069290013N},
  \DOIprefix\doi{https://doi.org/10.1016/0261-5606(92)90013-N}.
%Type = Article
\bibitem[{Margrabe(1978)}]{Margrabe1978}
\bibinfo{author}{Margrabe, W.}, \bibinfo{year}{1978}.
\newblock \bibinfo{title}{The value of an option to exchange one asset for
  another}.
\newblock \bibinfo{journal}{Journal of Finance} \bibinfo{volume}{33},
  \bibinfo{pages}{177--86}.
\newblock \DOIprefix\doi{10.1111/j.1540-6261.1978.tb03397.x}.
%Type = Article
\bibitem[{Milevsky and Posner(1998)}]{Milevsky}
\bibinfo{author}{Milevsky, M.A.}, \bibinfo{author}{Posner, S.E.},
  \bibinfo{year}{1998}.
\newblock \bibinfo{title}{A closed-form approximation for valuing basket
  options}.
\newblock \bibinfo{journal}{The Journal of Derivatives} \bibinfo{volume}{5},
  \bibinfo{pages}{54--61}.
\newblock \URLprefix \url{https://jod.pm-research.com/content/5/4/54},
  \DOIprefix\doi{10.3905/jod.1998.408005},
  \href{http://arxiv.org/abs/https://jod.pm-research.com/content/5/4/54.full.pdf}{\tt
  arXiv:https://jod.pm-research.com/content/5/4/54.full.pdf}.
%Type = Inproceedings
\bibitem[{Nair and Hinton(2010)}]{relu}
\bibinfo{author}{Nair, V.}, \bibinfo{author}{Hinton, G.E.},
  \bibinfo{year}{2010}.
\newblock \bibinfo{title}{Rectified linear units improve restricted boltzmann
  machines}, in: \bibinfo{booktitle}{Proceedings of the 27th International
  Conference on International Conference on Machine Learning},
  \bibinfo{publisher}{Omnipress}, \bibinfo{address}{Madison, WI, USA}. p.
  \bibinfo{pages}{807–814}.
%Type = Book
\bibitem[{Nocedal and Wright(2006)}]{Nocedal}
\bibinfo{author}{Nocedal, J.}, \bibinfo{author}{Wright, S.J.},
  \bibinfo{year}{2006}.
\newblock \bibinfo{title}{Numerical Optimization}.
\newblock \bibinfo{edition}{second} ed., \bibinfo{publisher}{Springer},
  \bibinfo{address}{New York, NY, USA}.
%Type = Article
\bibitem[{Raissi(2018)}]{raissi2018}
\bibinfo{author}{Raissi, M.}, \bibinfo{year}{2018}.
\newblock \bibinfo{title}{Forward-backward stochastic neural networks: Deep
  learning of high-dimensional partial differential equations}.
\newblock \bibinfo{journal}{arXiv preprint arXiv:1804.07010} .
%Type = Article
\bibitem[{Ramsundar et~al.(2015)Ramsundar, Kearnes, Riley, Webster, Konerding
  and Pande}]{Ramsundar}
\bibinfo{author}{Ramsundar, B.}, \bibinfo{author}{Kearnes, S.M.},
  \bibinfo{author}{Riley, P.}, \bibinfo{author}{Webster, D.},
  \bibinfo{author}{Konerding, D.E.}, \bibinfo{author}{Pande, V.S.},
  \bibinfo{year}{2015}.
\newblock \bibinfo{title}{Massively multitask networks for drug discovery}.
\newblock \bibinfo{journal}{ArXiv} \bibinfo{volume}{abs/1502.02072}.
%Type = Book
\bibitem[{Shreve(2004)}]{Shreve2004}
\bibinfo{author}{Shreve, S.E.}, \bibinfo{year}{2004}.
\newblock \bibinfo{title}{Stochastic calculus for finance 2, Continuous-time
  models}.
\newblock \bibinfo{publisher}{Springer}, \bibinfo{address}{New York, NY;
  Heidelberg}.
\newblock \URLprefix
  \url{http://www.worldcat.org/search?qt=worldcat_org_all&q=0387401016}.
%Type = Article
\bibitem[{Sirignano and Spiliopoulos(2018)}]{Sirignano2018}
\bibinfo{author}{Sirignano, J.}, \bibinfo{author}{Spiliopoulos, K.},
  \bibinfo{year}{2018}.
\newblock \bibinfo{title}{Dgm: A deep learning algorithm for solving partial
  differential equations}.
\newblock \bibinfo{journal}{Journal of Computational Physics}
  \bibinfo{volume}{375}, \bibinfo{pages}{1339--1364}.
\newblock \DOIprefix\doi{10.1016/j.jcp.2018.08.029}.

\end{thebibliography}

\end{document}